# Towards a Common Testing Terminology for Software Engineering and Data Science Experts

Lisa Jöckel[1], Thomas Bauer[1], Michael Kläs[1], Marc P. Hauer[2], Janek Groß[1]

[1] Fraunhofer Institute for Experimental Software Engineering IESE,
Fraunhofer-Platz 1, 67663 Kaiserslautern, Germany
[2] Algorithm Accountability Lab, TU Kaiserslautern, Gottlieb-Daimler-Strasse 48,
67663 Kaiserslautern, Germany
{lisa.joeckel, thomas.bauer, michael.klaes,
janek.gross}@iese.fraunhofer.de, hauer@cs.uni-kl.de

**Abstract.** Analytical quality assurance, especially testing, is an integral part of software-intensive system development. With the increased usage of Artificial Intelligence (AI) and Machine Learning (ML) as part of such systems, this becomes more difficult as well-understood software testing approaches cannot be applied directly to the AI-enabled parts of the system. The required adaptation of classical testing approaches and the development of new concepts for AI would benefit from a deeper understanding and exchange between AI and software engineering experts. We see the different terminologies used in the two communities as a major obstacle on this way. As we consider a mutual understanding of the testing terminology a key, this paper contributes a mapping between the most important concepts from classical software testing and AI testing. In the mapping, we highlight differences in the relevance and naming of the mapped concepts.

## 1 Motivation

In complex software-intensive systems, analytical quality assurance (QA) activities, especially software testing, have proven to be crucial for achieving high product quality. Due to the increasing relevance of Artificial Intelligence (AI) and Machine Learning (ML) as part of software systems, the question arises how AI/ML-enabled systems, and especially their AI/ML-based components, should be tested. The functionality of such components, which we refer to as *data-driven components* (DDCs), is not explicitly defined by a specification and implemented by a programmer within the code. Instead, it is given by a – usually complex and not human-understandable – model that is automatically derived from a data sample via a learning algorithm. Due to properties such as limited specification and understandability, the transfer of classical test approaches is not trivial.

In the field of AI, the QA of DDCs has so far played a minor role and has mainly been done by applying specific evaluation criteria such as accuracy to a previously unseen subset of the available data. As the application of AI is being extended to ever more domains, including safety-critical areas such as autonomous driving, industrial automation, or medical applications, the demand for QA has also increased in recent years. New techniques are being proposed and quality aspects like fairness, robustness, and explainability are becoming more important. Although some approaches for testing DDCs are described in the literature [1] including some very sophisticated ones, their relation to classical software testing and system QA is not covered sufficiently yet.

We see the potential to exploit experiences and concepts from the field of classical software testing for the QA of AI-based systems and components. To this end, collaboration and direct exchange between experts from both fields are important. This is, however, impeded by different terminologies and meaning of terms, which leads to misunderstandings and makes it more difficult to relate to work from the respective other field.

*Contribution*: In this paper, we make a first step towards a common terminology. We use established terms from classical software testing as a basis to map corresponding concepts from the field of AI to it, pointing out differences and key challenges in transferring known concepts. The proposed mapping was developed in an interdisciplinary collaboration among the authors, who have many years of experience in at least one of the two fields, partly in both. We intend this to be a stimulus and a basis for discussions aimed at building a common understanding between experts of both fields.

In Section 2, we will describe some background regarding DDCs. Section 3 provides an overview of related work on testing terminology. Section 4 presents a mapping between testing terminology for classical software and AI. Section 5 concludes the paper.

## 2  Background on Data-Driven Components

In this section, we provide some background on DDCs that is relevant for understanding the discussions on the test concepts in Section 4. To this end, we will briefly describe a typical DDC lifecycle as well as supervised learning, and introduce an example use case.

As QA is done throughout the lifecycle of a DDC, we use an adapted lifecycle for DDCs [2] that allows differentiating the purposes of QA measures and datasets (see Fig. 1). Multiple datasets are needed for different purposes (e.g., training, validation, testing) during the DDC lifecycle. As the functionality of DDCs is derived from and evaluated on data, this is a key aspect that needs to be treated with caution. In the DDC lifecycle, the *specification* defines, among other things, the task of the AI, its target application scope (TAS) [3], and its required quality characteristics. The TAS is related to the operational design domain in the automotive domain. It defines in which context and under which conditions the DDC is considered applicable; hence, it is an important building block for testing and needs to be reflected by the test dataset. During *construction*, the data-driven model (DDM) is built as core of the DDC. Its input-outcome relationship is derived from a data sample, i.e., a training dataset composed for the intended task. The expected behavior of DDCs is therefore only specified for a subset of all possible input data. For previously unseen inputs, the expected behavior cannot be fully assured. We distinguish two phases of analytical QA activities during design time according to their purposes: (1) *Analysis* activities aim at finding potential weak points to improve the DDM, like explainability approaches. The results from the analysis are fed back to the construction phase. (2) T*esting* activities aim at providing quantitative evidence for the specified requirements, which are generated on a test dataset that is representative for the TAS. This differentiation into analysis and testing is a distinct feature of the lifecycle of DDCs, as eliminating faults based on incorrect outcomes is difficult [4]. The analysis and testing phases take place before the AI component is deployed. During *operation*, monitoring activities are needed to ensure that the application is in line with the specification. In the remainder of this paper, our focus will be on *analytical QA activities in the testing phase*.

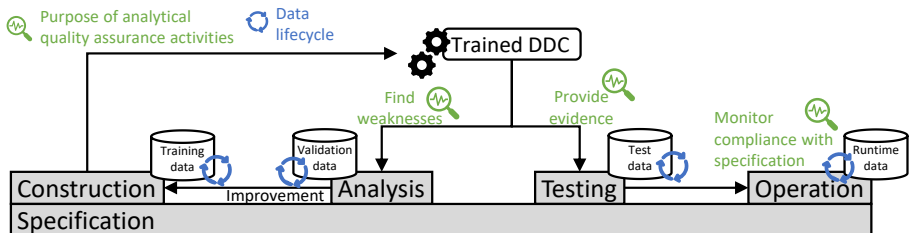

**Fig. 1.** Lifecycle model of a DDC with analytical quality assurance for different purposes.

Techniques for building DDCs can be grouped by the degree of supervision they need, which in turn influences the possibilities and raises different challenges for testing. Our focus is on *DDCs using supervised learning techniques*, where there is ground truth information for the outcomes, i.e., each data point is labeled with its expected outcome.

We will later refer to an example DDC whose task is traffic sign recognition (TSR), i.e., classification of the traffic sign type on a given input image, on German roads.

## 3 Related Work on Testing Terminology

Software testing has been a fundamental discipline in software engineering since the very beginning. Therefore, processes, terms, and definitions for software testing have been defined since the 1980s, leading to standards such as IEEE 829 for Software Test Documentation [5], and the IEEE 610 Standard Computer Dictionary [6], which still represent the basis for fundamental terms and definitions in software testing. They have been updated step by step and have been tailored for new domains and system classes [7], as well as being supplemented with new concepts, e.g., test coverage [8].

In contrast, the testing of AI-based software systems has only gained importance in recent years [4]. As there are many challenges related to the testing of AI [9, 4], a transfer of concepts and the corresponding terminology from classical software testing is not trivial. Lenarduzzi et al. provide a mapping between terms that are misleading or used differently in software engineering and AI [10]. Some works provide an overview of what has been done so far in transferring testing concepts, including the definition and relations of some testing terms [1, 4, 9]. These terms include, e.g., test input generation, adequacy criteria, oracle, testing level, online and offline learning. However, the number of considered terms is rather selective and not clearly oriented on the workflows for software and AI testing, which would improve relating the terminology of both fields to establish a common understanding. To the best of our knowledge, a comprehensive mapping between the terminologies – considering differences and common aspects as well as their relations to the testing workflows – has not been performed yet.

## 4 Mapping of Software and AI Testing Terminology

In this section, we will first provide an overview of the basic workflow and terminology in classical software testing. Then we will relate common concepts and terminology from the field of AI testing to them, pointing out some difficulties in doing so. A mapping of the testing workflows is illustrated in Fig. 2, including testing terms, instances for the terms from example components calculating the next gas station or classifying traffic sign images, and highlighted differences in the workflows, i.e., different concepts and terms or variations in their importance. Each of the following subsections describes a part of the workflow.

### 4.1 Test Abstraction Levels and Objects

**Software Testing.** In software engineering, *testing* is defined as "an analytical QA activity in which systems, subsystems, or components are executed under specified conditions, the results are observed or recorded, and an evaluation is made of some aspect of the system or component" [6]. This means that testing is performed on specific abstraction levels (component, integration, system) [13] when executable artifacts such as program code or executable models become available as test objects. A *test object* or test item is defined as "a software or system item that is an object of testing" [5] and implements a (sometimes implicit) specification. A *specification* is "a document that specifies, in a complete, precise, verifiable manner, the requirements, design, behavior, or other characteristics of a system or component, and, often, the procedures for determining whether these provisions have been satisfied" [6]. The test object is tested against the requirements, i.e., the required capabilities of the system or system component [6], and against quality characteristics. In this work, we focus on software *component testing*, which is defined as "testing of individual hardware or software components or groups of related components" where a *component* is "one of the parts that make up a system […] and may be subdivided into other components" [6]. Each component contributes to a specific function or set of functions of its associated system.

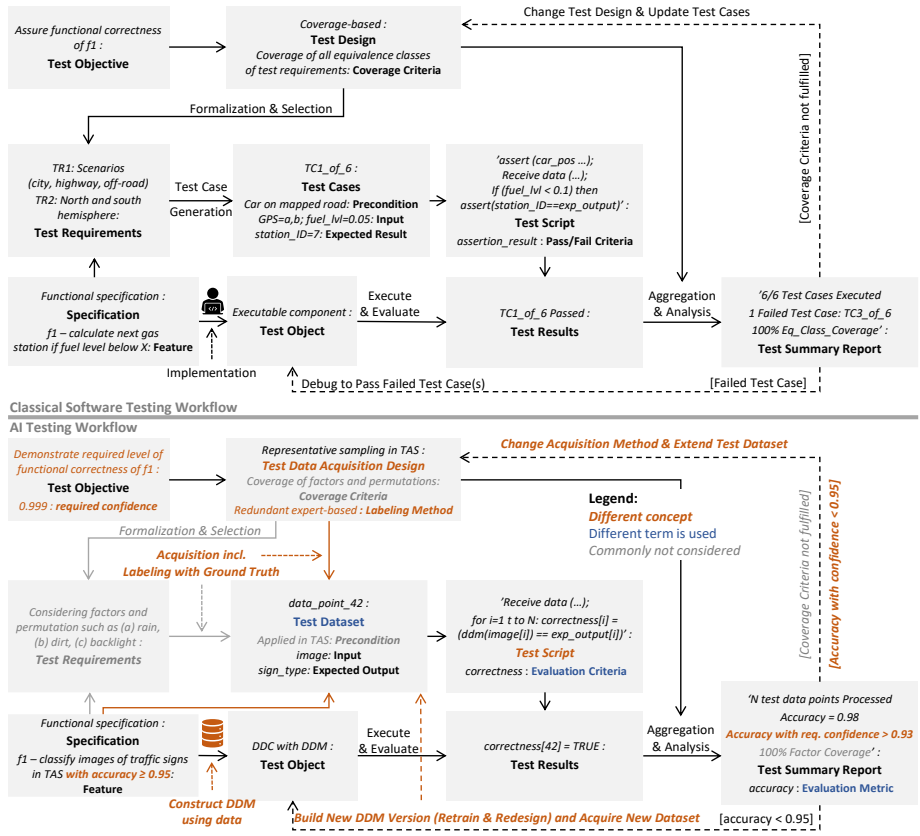

**Fig. 2**. Comparison of testing workflows for classical software (top) and AI (bottom).

**AI Testing.** We consider DDCs to be a counterpart to classical software components. A DDC may consist of sub-components that are organized in pipelines that include some data pre- and post-processing in addition to the trained DDM [14]. Since data pre- and post-processing can be addressed with software testing approaches, AI testing focuses on the DDM. As isolated testing of the implemented training algorithm does not reveal whether the trained model successfully derived the intended behavior from the training data, the trained DDM is considered to be the *test object*. Yet, as the behavior of the DDM is learned from data, ensuring that the data itself meets certain requirements becomes increasingly important. Although data is not an 'executable artifact' on its own, but only in combination with the model, certain characteristics of the dataset can be checked (e.g., inclusion of edge cases) with regard to the intended task of the DDC and its TAS. Contrary to classical software components, the behavior of the DDC cannot be described in a complete and verifiable manner as part of the *specification* as its functionality is not defined by the developer but is derived from data. For testing, functional correctness is mostly regarded as a *quality characteristic* (others might be fairness, robustness, and explainability). However, unlike in software testing, requirements on functional correctness need to be given a probabilistic sense (e.g., stop signs are correctly detected with a probability of 91%) as the input-outcome relationship cannot be fully specified and uncertainty in the DDC outcomes cannot be fully eliminated. For integration- and system-level tests, aspects beyond the scope of this paper need to be considered when a DDC is involved, like processing possibly incorrect DDC outcomes in other system components.

### 4.2 Getting from Test Objective to Test Cases

**Software Testing.** A *test objective* is defined as "an identified set of software features to be measured under specific conditions by comparing the actual behavior with the required behavior described in the documentation or specification of the test object" [5]. Based on this, the *test design* describes the method used to systematically formalize and select test requirements, where a *test requirement* is defined as "a specific element of an artifact (such as the functional system specification) that a test case or a set of test cases must cover or an artifact property that the test case or the test case set must satisfy" [12]. A *test case* is "a set of input values, execution conditions, and expected results developed for a particular objective, such as to exercise a particular program path or to verify compliance with a specific requirement" [6]. The quality and completeness of test cases are assessed by *test coverage criteria*, which define the selection rules for determining or collecting a set of test requirements to be considered [12]. The actual *test coverage* is defined as "the degree to which a test case or set of test cases addresses all selected test requirements of a given test object" [6]. The degree is usually expressed as a percentage. Test coverage is often used as an acceptance and stopping criterion for specifying test cases [8].

**AI Testing.** The *test objective* is commonly to show a required level of functional correctness as defined in the specification, e.g., an accuracy of at least 95%. As functional correctness is measured on a data sample, we can additionally require a certain confidence in the evaluation, e.g., requiring a confidence of 99.9% that the actual accuracy is not lower than 95%. This way, we reduce the risk of wrongly assuming an accuracy level that is too high.

In general, DDCs have to be tested on data that was not used during the development of the DDC, i.e., the *test dataset*, which also contains ground truth information for supervised models. Each data point can be seen as a *test case* providing the model input and the expected outcome, e.g., an image showing a stop sign as model input with the corresponding sign type as expected outcome. Execution preconditions are usually not defined explicitly, but implicitly, as the inputs should be collected from the TAS. Determining the expected outcome, i.e., the ground truth, is more difficult in most cases than for classical software components as the labeling is mainly done manually, not always unambiguous, and sometimes involves the observation of complex empirical processes, e.g., when we need to determine whether a certain cancer therapy was successful. This limits the amount of data available and the freedom in designing test cases. Sometimes, this issue is addressed by simulations to generate labeled synthetic data or data augmentations to add changes to a data point in a way that the ground truth is still known [15]. However, due to limitations regarding the realism of such data, it is not clear to which degree the testing performance can be transferred to real inputs during operation. Commonly, the *test dataset is acquired* by a representative sample for the TAS (without defining test requirements). The method for labeling the data with ground truth information is also part of the data acquisition.

In analogy to classical software testing, *test requirements* could be defined. For the example DDC, this could be done by considering relevant factors influencing the input data quality, e.g., rain or a dirty camera lens. Here, *test coverage criteria* would be based on the influence factors and their permutations. However, defining test requirements in this way involves expert knowledge and is often not done explicitly in practice, which potentially leads to important influence factors not being (sufficiently) considered in the data, such as snow-covered traffic signs. Other possible coverage criteria are related to code coverage in classical software, like neuron coverage for neural networks demanding neurons to exceed a defined activation level [16]. Coverage criteria are often difficult to transfer to DDCs as they usually operate in an open context with many unforeseeable situations. Additionally, small changes in the input might lead to large variations in the outcome [17]. Therefore, the stopping criterion for testing is mostly handled trivially by stopping when all data points in a given test dataset have been processed. However, this does not necessarily reveal to which extent the test objective is addressed.

### 4.3 Test Execution and Evaluation

**Software Testing.** For the test execution, specific *test scripts* have to be derived from the test cases to enable a connection to the execution environment and the test tools, to stimulate the test object with concrete signals, messages, as well as function calls, and to record the system responses for the subsequent evaluation [13]. The actual response is compared with the expected response defined in the test case and implemented in the test script to determine the *test result*, i.e., whether or not a specific test case has passed or failed [7]. The *test summary report* includes a summary of the test activities and results, considering failed test cases and achieved coverage level [5]. For the failed test cases, the underlying faults are localized and fixed to improve the test object. Insufficient coverage leads to changes of the test design and, hence, an updated set of test cases.

**AI Testing.** *Test scripts* in the sense of software testing do not play a prominent role in AI testing. The reason is that DDCs are commonly stateless software components with well-defined, simple interfaces (e.g., taking as input an image of a defined size and providing as outcome a sign type). Thus, there is no need for individual scripts implementing different test cases but just for a single script that loads the test dataset, executes the DDC with each input, and then computes the *test results* applying the evaluation criteria, e.g., correctness of the DDC outcome, on each pair of obtained/expected outcomes. AI *test reports* commonly focus on aggregated results for the relevant evaluation metric, e.g., accuracy, without indicating which test cases failed. The reason is that unlike in software testing, where faults causing a specific failure can be localized and fixed in the test object individually, no concept equivalent to a fault exists for DDMs. The DDC will thus only be revised if the test report indicates that the test objective is not met. In such cases, a new DDM has to be constructed and a new test dataset needs to be acquired to avoid the situation that the construction of the new DDM can make use of knowledge about the test data to be applied later. The test objective might require that the evaluation metric result is met with a given confidence. If the test report indicates that the uncertainty in the evaluation metric result is too high, i.e., the evaluation metric result with confidence is lower than the required evaluation metric result, the test dataset may be extended by acquiring additional test cases, thereby reducing the uncertainty in the evaluation metric result.

## 5 Conclusion

In classical software testing, well-elaborated test concepts and processes exist. Due to the different nature of DDCs, transferring known test concepts to AI is not trivial and their applicability is not easy to assess. Therefore, we propose intensifying the exchange of experience between experts from both communities. In this paper, we contribute to this by encouraging discussions on mapping terminology from software testing to AI.

We focused on supervised DDCs, well aware that unsupervised or reinforcement learning might raise further challenges, e.g., no ground truth information being available. Furthermore, the benefits of AI testing vary from classical software testing. While insufficient or incorrect behavior of the DDC might be revealed, this rarely provides information on how the behavior came to happen (e.g., due to the model hyperparameters, insufficient training data, or the training process) and thus how to improve the DDC. Additionally, we only have only a partial specification for DDCs based on a data sample, and therefore some uncertainty always remains in the outcomes. This raises the question of how test evidences need to be interpreted and what this implies in relation to classical test evidences and the expected runtime performance; i.e., the validity of the test results cannot be guaranteed for situations outside the TAS, which are often difficult to detect.

**Acknowledgments.** Parts of this work have been funded by the Observatory for Artificial Intelligence in Work and Society (KIO) of the Denkfabrik Digitale Arbeitsgesellschaft in the project "KI Testing & Auditing" and by the project "AIControl" as part of the internal funding program "KMU akut" of the Fraunhofer-Gesellschaft.